\documentclass[preprint,aps]{revtex4}
\usepackage{amsmath}
\usepackage{amstext}
\usepackage{amssymb}
\usepackage{graphicx}
\usepackage{natbib}

\def\eq#1{Eq.~(\ref{#1})}
\def\fig#1{Fig.~\ref{#1}}

\begin{document}

\title{Focal adhesions as mechanosensors: the two-spring model}
\author{Ulrich S. Schwarz}
\author{Thorsten Erdmann}
\author{Ilka B. Bischofs}
\affiliation{Interdisciplinary Center for Scientific Computing, University of Heidelberg, 
Im Neuenheimer Feld 368, D-69120 Heidelberg, Germany}
\affiliation{Max Planck Institute of Colloids and Interfaces, 14424 Potsdam, Germany}

\begin{abstract}
  Adhesion-dependent cells actively sense the mechanical properties of
  their environment through mechanotransductory processes at focal
  adhesions, which are integrin-based contacts connecting the
  extracellular matrix to the cytoskeleton.  Here we present first
  steps towards a quantitative understanding of focal adhesions as
  mechanosensors. It has been shown experimentally that high levels of
  force are related to growth of and signaling at focal adhesions.  In
  particular, activation of the small GTPase Rho through focal
  adhesions leads to the formation of stress fibers. Here we discuss
  one way in which force might regulate the internal state of focal
  adhesions, namely by modulating the internal rupture dynamics of
  focal adhesions. A simple two-spring model shows that the stiffer
  the environment, the more efficient cellular force is built up at
  focal adhesions by molecular motors interacting with the actin
  filaments.
\end{abstract}

\maketitle

\section{Introduction}

During recent years, tremendous progress has been made in regard to a
quantitative understanding of the metabolic, signal transduction and
genetic networks characteristic of biological systems
\cite{s:kita02,s:alon03,s:alm03}.  Although network approaches capture
many of the essential aspects of simple organisms, for higher
organisms a quantitative and systems-level understanding also has to
include structural aspects, including the spatial organisation and
mechanical properties of cells. In particular, modelling tissues and
organs requires a quantitative understanding of the roles played by
cytoskeleton, membranes, and the extracellular matrix (ECM).

One field which cannot be understood completely without considering
biochemical and structural aspects on an equal footing is cell
adhesion, which is an essential element of many physiological
situations, including development, tissue maintenance, wound healing,
angiogenesis, and cell migration \cite{c:gumb96}. In general, most
cell types require anchorage to the ECM to proliferate.  Moreover,
cell adhesion also determines how cells interpret soluble ligands like
hormones and growth factors
\cite{c:stup02,c:guo04}. The behaviour of adhering cells is strongly
influenced by the chemical, topographical and mechanical properties of
the surfaces they attach to \cite{c:curt01}. During recent years,
experiments with elastic substrates have shown that elastic properties
of the extracellular evironment are also highly relevant for cellular
decision making \cite{c:pelh97,c:lo00,c:wong03,c:engl04,c:geor05}.

A growing body of evidence now suggests that the essential link
between the mechanical properties of the extracellular environment and
cellular decision making are mechanotransductory processes at
integrin-based cell-matrix contacts
\cite{c:chic98,c:galb98,c:geig01a,c:kats04}. For cells spreading on
flat substrates, cell-matrix contacts initially form as focal
complexes close to the lamellipodium.  Depending on the presence of
appropriate signals, focal complexes can mature into focal adhesions
which are connected to actin stress fibers. Focal adhesions have a
twofold purpose. As they connect the actin cytoskeleton with the ECM,
they guarantee structural integrity. Equally important, they are also
strong signaling centers. In fact more than 50 different kinds of
proteins are known to localize to the cytoplasmic plaque of focal
adhesions, many of which are known signaling molecules. Therefore
focal adhesions provide an excellent opportunity to study the
interplay between biochemical and structural aspects in biological
systems.

The details of the mechanosensory processes at focal adhesions are
still elusive. It has been shown some time ago that application of
force on integrin-based contacts between cells and ligand-coated beads
leads to contact reinforcement and mechanotransduction
\cite{c:wang93,c:choq97}. Recently, force reconstruction at single
focal adhesions on compliant substrates showed that the internal
forces exerted at focal adhesions correlate with their sizes
\cite{uss:bala01,c:tan03}. In a complementary study, it was shown that
force exerted externally by a micropipette leads to growth of those
focal adhesions which are tensed \cite{uss:rive01}.  Other recent
experiments imply both a membrane-independent stretch response of the
protein network connected to focal adhesions \cite{c:sawa02} as well
as some role for stretch-activated ion channels \cite{c:mune04a}. In
fact it is very likely that several force-mediated mechanisms work in
parallel at focal adhesions, including changes in integrin and
extracellular ligand densities, rearrangements in the cytoplasmic
plaque, stretch-activated ion channels and opening of cryptic binding
sites in focal adhesion molecules \cite{c:bers03}. Recently a
quantitative model has been introduced which explains anisotropic
growth of focal adhesions under force by density variations in the
sheared layer of integrins \cite{c:nico04a,c:nico04b}. Other
theoretical efforts have modelled force-mediated growth as strain
relaxation due to incorporation of new material, phase transitions due
to force-mediated coupling between neighboring receptors and
force-mediated release of a soluble signal.  However, a systems-level
description of focal adhesions as mechanosensors has not been
presented yet.

In this contribution, we discuss several modelling efforts which in
the future might be integrated into such a systems-level understanding
of focal adhesions.  Such a description will have to integrate the
effects of extracellular elasticity, molecular motor activity, and
signal transduction. We start with a discussion of integrin signaling
at focal adhesion and how it relates to the spatial and temporal
organization of cells. Next we describe a simple model for the
stochastic rupture dynamics of adhesion clusters under force, which
quantitatively demonstrates that the internal state of adhesion
clusters can be regulated by force. Finally we introduce a new model
(\textit{two-spring model}), which shows in a quantitative way how
extracellular elasticity might modulate the build-up of intracellular
force at focal adhesions.

\section{Integrin signaling at focal adhesions}

Although physical concepts like force and elasticity are essential to
understand active mechanosensing at focal adhesions, the biochemical
aspects of these systems are equally important and far from
understood.  Focal adhesions are based on heterodimeric
transmembrane-receptors from the integrin family, which connect the
ECM with the actin cytoskeleton. Integrins are large allosteric
machines which are activated both by biochemical and mechanical cues
and which transmit both inside-out and outside-in signals
\cite{c:hyne02}. For mammals, 24 integrin variants are known, which
bind to different subsets of ECM-ligands. For example, the main
integrin-receptors for fibronectin and vitronectin are $\alpha_5
\beta_1$ and $\alpha_{\nu} \beta_{3}$, respectively. Interestingly,
cancer cells switch their integrins: they loose integrins like
$\alpha_3 \beta_1$, which mediate adhesion, and upregulate integrins
like $\alpha_{\nu} \beta_{3}$, which promote migration and survival in
new environments \cite{c:guo04}. The whole complexity of the integrin
systems becomes apparent when one considers the interaction with the
the cytoplasmic plaque and the signaling to the cytoskeleton
\cite{c:geig01a}.

\begin{figure}
\begin{center}
\includegraphics[width=1.0\textwidth]{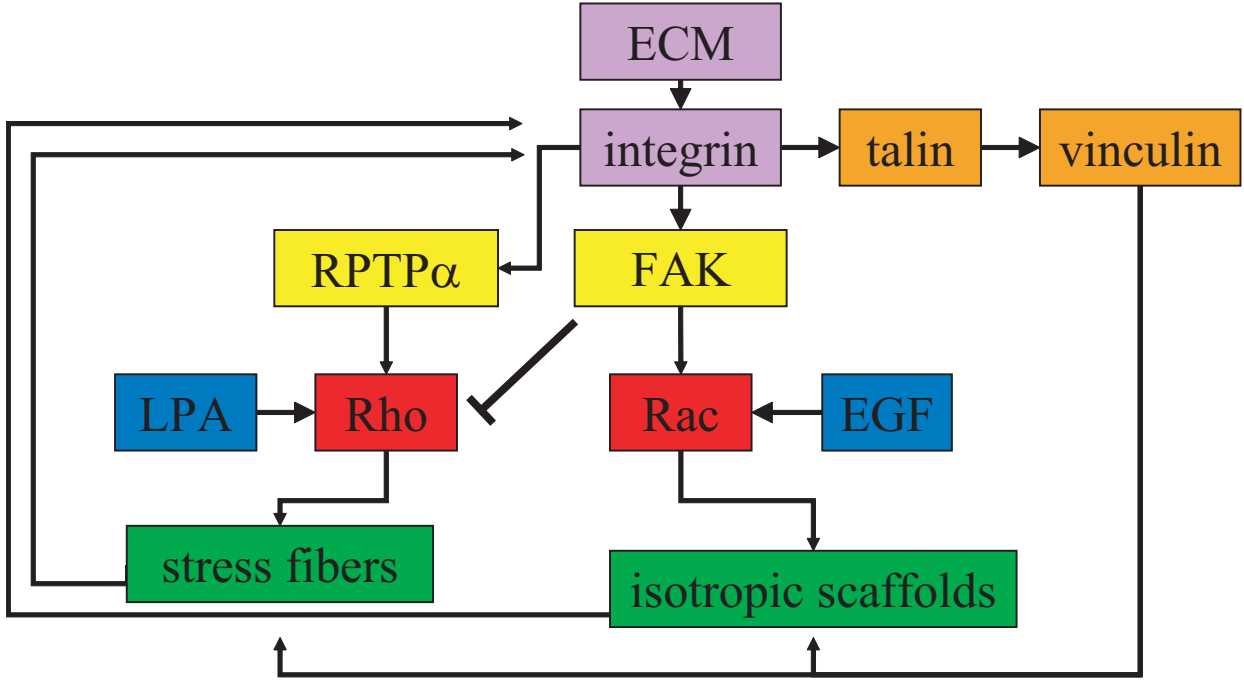}
\caption{Focal adhesions are tightly regulated by signaling events.
  Important downstream targets for integrins include FAK,
  RPTP-$\alpha$ and talin. Enzymatically active molecules like FAK and
  RPTP-$\alpha$ lead to activation of the small GTPases Rac and Rho,
  which regulate the structural organization of the actin
  cytoskeleton. This in turn feeds back to nucleation and growth of
  focal adhesions. FAK-activation through integrin ligation also leads
  to transient downregulation of Rho, resulting in an antagonistic
  role of Rac and Rho. The talin-mediated link between integrins and
  actin is stabilized by vinculin, but as both proteins can exist in
  closed and open conformations, they might also play a more active
  role in mechanosensation at focal adhesions.}
\label{Signaling}
\end{center}
\end{figure}

In \fig{Signaling}, we schematically depict some of the aspects which
are known to be important in this regard. The scheme deliberately
focuses on three important downstream targets of integrin signaling to
the actin cytoskeleton \cite{c:dema}.  Focal adhesion kinase (FAK) is
a protein tyrosine kinase which has been shown to be a key component
of mechanosensing at focal adhesions \cite{c:wang01}. It is activated
by integrin ligation and one of its main downstream targets is the
small GTPase Rac, which leads to reorganization of the actin
cytoskeleton into an isotropic network structure. At the same time,
FAK-activation downregulates another small GTPase, Rho, mainly through
activation of p190RhoGAP. Rho promotes the reorganization of the actin
cytoskeleton into stress fibers and it often has an antagonistic role
to Rac. Both small GTPases belong to the Rho-family and are also
activated by pathways involved in cell survivial (epidermal growth
factor (EGF) and lysophosphatidic acid (LPA) in the cases of Rac and
Rho, respectively). Upregulation of Rac and downregulation of Rho is
typcial for phases of spreading and cell migration, when focal
complexes and lamellipodia are more prominent than focal adhesions and
stress fibers.  However, the initial dip in Rho-activity is often
followed by long-term activation, albeit in a ligand-specific and
cell-type-specific manner \cite{c:bers03}. This typically corresponds
to the phase of mature adhesion, which is discussed here. Although
experimental findings are conflicting, there is good evidence that the
receptor-like protein tyrosine phosphatase RPTP-$\alpha$ activates Rho
through the tyrosine kinase Fyn and a RhoGEF which has not been
identified yet \cite{c:wich03}. Irrespective of the detailed
mechanism, Rho-activation has been shown to be an essential part for
the force-mediated stabilization of focal adhesions \cite{uss:rive01}.
The main issue here is that Rho-mediated activation of myosin II
molecular motor activity as well as formation of stress fibers is
essential for maturation of focal adhesions, thus providing positive
feedback to growing adhesions. Rac-mediated organization of the actin
cytoskeleton into isotropic networks might provide positive feedback
for the growth of focal complexes, but possibly in a force-independent
way.

A third major player in focal adhesions is talin, one of the four
proteins known to link the integrins directly with the actin
cytoskeleton. Talin is essential for early focal adhesion
reinforcement under force \cite{c:jian03} and leads to recruitment of
vinculin, which also stabilizes focal adhesions. Both talin and
vinculin can exist in closed and open conformations, a fact which
might be related to the mechanosensor at focal adhesions
\cite{c:bers03}.  They also might act as nucleators for the actin
cytoskeleton, thus locally modulating the effects of the small GTPases
Rac and Rho. Finally it is interesting to note that the actin
cytoskeleton also features crosstalk to the microtubule system.  For
example, it has been shown that one of the main downstream targets of
Rho is mDia \cite{uss:rive01}, which might regulate microtubule
polymerization. Moreover it has been found that microtubules are
targeted into mature focal adhesions, possibly in order to deliver
some kind of death signal \cite{c:kryl03}.

The scheme presented in \fig{Signaling} shows that there exists a
positive feedback involving integrin ligation, assembly of the
cytoplasmic plaque, Rho- and Rac-signaling to the cytoskeleton and
reorganization of the cytoskeleton. In the case of Rho-signaling, an
essential element of this feedback is generation of stress through
myosin II molecular motors and growth of focal adhesions under force.
One of the future challenges in this field is a more complete and
data-based description of the interplay between signaling at and
spatial organization of integrin-based adhesions and the actin
cytoskeleton.  In order to understand the role of force in the
feedback loop between integrins and actin cytoskeleton, physical
mechanisms have to be identified by which force affects the state of
focal adhesions.

\section{Rupture dynamics of adhesion clusters under force}

\begin{figure}
\begin{center}
\includegraphics[width=1.0\textwidth]{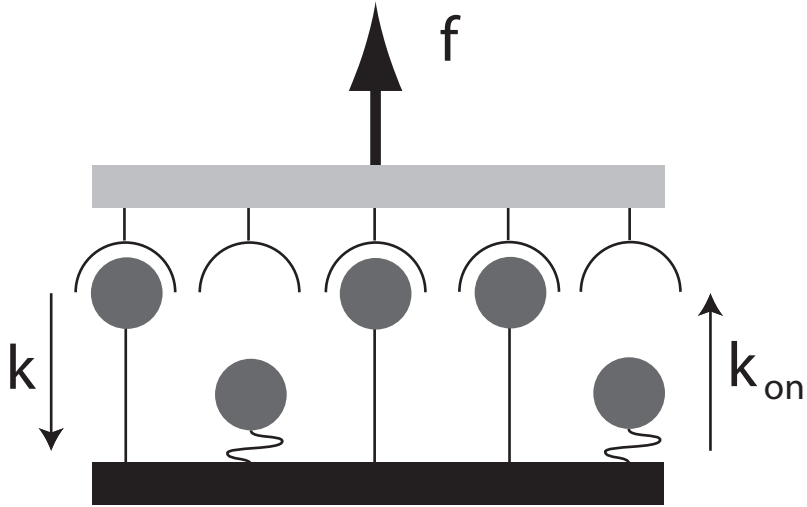}
\caption{Schematic representation of an adhesion cluster
under force. Closed bonds rupture with a force-dependent rate and open
bonds close with a force-independent rebinding rate. Non-trivial
cooperativity results when force is shared between closed bonds.}
\label{PRLThorsten1}
\end{center}
\end{figure}

In order to study how force affects adhesion clusters in general, we
recently studied a simple model for the stochastic dynamics of
parallel bonds under shared constant loading
\cite{uss:erdm04a,uss:erdm04c}. This model is a stochastic version of
a classical yet deterministic model which has been introduced by Bell
\cite{c:bell78}. The model assumes that $N_t$ receptor-ligand bonds
have been clustered on opposing surfaces, of which the upper one acts
as a rigid transducer which transmits the constant force $F$
homogeneously onto the array of bonds. In our model, $N_t$ is a
constant, but in future work it might be combined with a growth model
for adhesion clusters \cite{c:nico04a,c:nico04b}. At each time, $i$
($0 \le i \le N_t$) bonds are closed and $N_t - i$ bonds are open.
Closed bonds are assumed to rupture with a force-dependent rupture
rate $k = k_0 e^{F / (i F_b)}$, where $k_0$ is the unstressed
(intrinsic) rupture rate (typically around 1/s) and $F_b$ the internal
force scale (typically a few pN) of the adhesion bonds. The
exponential dependance between force and rupture rate results from a
Kramers-type description of bond rupture as escape over a transition
state barrier \cite{c:evan97}. The factor $i$ results because force is
assumed to be shared equally between closed bonds, which holds true
when the transducer is connected to a soft spring (in the opposite
limit of a stiff spring, all bonds feel the same force and
cooperativity is lost). Open bonds are assumed to rebind with a
force-independent rebinding rate $k_{on}$. A schematic representation
of our model is shown in \fig{PRLThorsten1}. The model has three
dimensionless parameters, namely cluster size $N_t$, dimensionless
total force $f = F / F_b$ and dimensionless rebinding rate $\gamma =
k_{on} / k_0$. With dimensionless time $\tau = k_0 t$, it leads to the
following one-step master equation
\begin{equation} \label{MasterEquation}
\frac{dp_i}{d\tau} = r_{i+1} p_{i+1} + g_{i-1} p_{i-1} - [ r_i + g_i ] p_i
\end{equation}
where $p_i(\tau)$ is the probability that $i$ bonds are closed at time
$\tau$ and the $r_i$ and $g_i$ are the reverse and forward rates
between the possible states $i$:
\begin{equation} \label{Rates}
r_i = r(i) = i e^{f/i}  \quad\text{and}\quad  g_i = g(i) = \gamma (N_t - i)\ .
\end{equation}
This equation implies $g_0 > 0$, that is, after rupture of the last
closed bond, new bonds are allowed to form. However, in many
situations of interest, rebinding from the completely dissociated
state is prevented by elastic recoil of the transducer. Therefore in
the following we use $g_0 = 0$ (absorbing boundary at $i = 0$). For
the mean number of closed bonds, $N(\tau) = \langle i \rangle$, one
can derive from \eq{MasterEquation}
\begin{equation} \label{FirstMoment}
\frac{dN}{d\tau} = \sum_{i=0}^{N_t} i \frac{dp_i}{d\tau}
       = - \langle r(i) \rangle + \langle g(i) \rangle\ .
\end{equation}
This suggests to study the following differential equation
\begin{equation} \label{DeterministicEquation}
\frac{dN}{d\tau} = \dot N = - r(\langle i\rangle) + g(\langle i\rangle) 
= - N e^{f / N} + \gamma (N_t - N) 
\end{equation}
as has been done by Bell \cite{c:bell78}.  However, this deterministic
treatment is a good approximation for the first moment of the
stochastic model only in the case of large systems. For small systems,
stochastic fluctuations in combination with the non-linearity and the
absorbing boundary lead to different results.

\begin{figure}
\begin{center}
\includegraphics[width=1.0\textwidth]{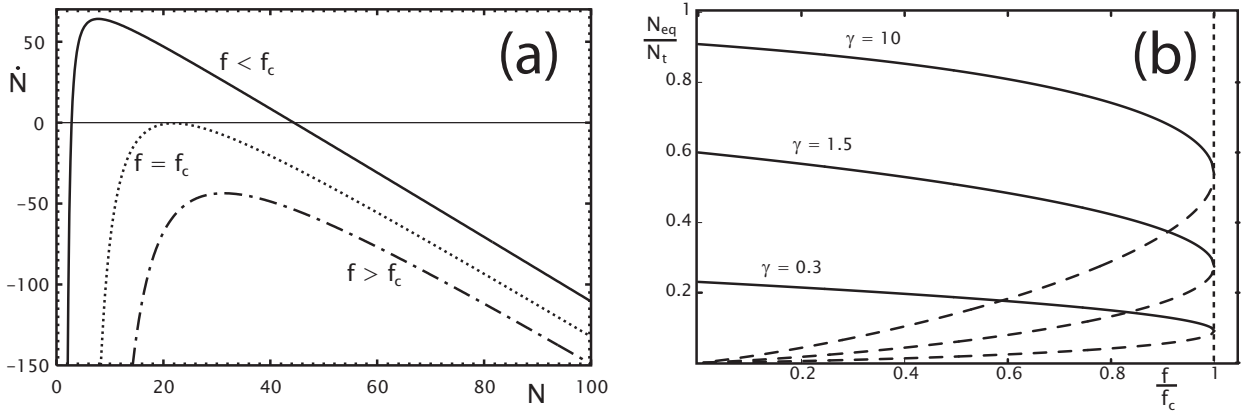}
\caption{Bifurcation analysis of the deterministic equation 
\eq{DeterministicEquation} for the number of closed bonds $N$.
(a) Below the critical force $f_c$, two fixpoints $N_{eq}$ exist
with $\dot N = 0$. The one with larger $N$ corresponds to a stable
state. (b) At the critical force $f_c$, the fixpoints vanish in a
saddle-node bifurcation. The exact values for the $N_{eq}$ depend on
the dimensionless rebinding rate $\gamma$.}
\label{bifurcation}
\end{center}
\end{figure}

While force destabilizes the cluster, rebinding stabilizes it.  We
first study this interplay in the framework of the deterministic
equation \eq{DeterministicEquation}. In \fig{bifurcation}a, we plot
$\dot N = dN / d\tau$ as a function of $N$ for several values of force
$f$.  This shows that two fixpoints $N_{eq}$ with $\dot N = 0$ exist
up to a critical force $f_c$, with the lower one being unstable (a
saddle) and the upper one being stable (a node). At $f = f_c$, the two
fixpoints collapse and stability vanishes in a saddle-node
bifurcation. The critical force can be calculated exactly to be
\begin{equation} \label{eq:fcrit}
f_c      = N_t\ \rm plog\left(\frac{\gamma}{e}\right)\ .
\end{equation}
Here the product logarithm $\rm plog(a)$ is defined as the solution
$x$ of $xe^x=a$. For $\gamma < 1$, we have $f_c \approx \gamma N_t /
e$. Thus the critical force vanishes with $\gamma$, because the
cluster decays by itself with no rebinding. For $\gamma > 1$, we have
$f_c \approx 0.5 N_t \ln \gamma$.  This weak dependence on $\gamma$
shows that the single bond force scale set by $F_b$ also determines
the force scale on which the cluster as a whole disintegrates.
\fig{bifurcation}b shows the full bifurcation diagrams for different 
values of the rebinding rate $\gamma$. The larger rebinding, the
larger are the values for the stable steady state. In particular, for
$f = 0$ we have $N_{eq} = \gamma N_t / (1+\gamma)$, that is $N_{eq}$
first increases linearly with $\gamma$ and then saturates towards
the maximal value $N_t$.

In conclusion, the bifurcation analysis of the Bell-model shows that
force can switch the stability of adhesion clusters. It is tempting to
speculate that focal adhesions might be regulated to be close to such
a critical state, because then small changes in cytoskeletal loading
would result in strongly accelerated cluster dynamics and larger
forces on single bonds. This in turn could trigger signaling events,
e.g.\ by exposure of cryptic binding sites. In fact the stress
constant at mature focal adhesions recently has been measured to be
around $\sim 5.5$ nN/$\mu m^2$ for different cell types and different
experimental conditions \cite{uss:bala01,c:tan03}. Using
\eq{eq:fcrit}, this idea can be used to estimate the rebinding rate in
focal adhesions, which has not been measured yet. Estimating $N_t =
10^4$ and using the dissocation parameters $k_0 = 0.012$ Hz and $F_b =
9$ pN for activated $\alpha_5 \beta_1$-integrin binding to fibronectin
\cite{c:li03} gives a rebinding rate $k_{on} = 0.002$ Hz.

Although the deterministic model gives non-trivial insight into
possible mechanisms for switching the state of focal adhesions by
force, it neglects fluctuation effects. In particular, cluster
lifetime is predicted to be infinite below the critical force $f_c$.
In the stochastic treatment, lifetime is finite for all parameter
values due to the possibility that the systems reaches the absorbing
boundary at the completely dissociated state.  Average cluster
lifetime $T$ then can be identified with the mean first passage time
to reach the state $i=0$, which can be calculated exactly from the
adjoint master equation. For one bond, one simply has $T = 1 / r_1 =
e^{-f}$, as suggested by Bell \cite{c:bell78}.  For two bonds, we find
\begin{equation} \label{eq:T_two_bonds}
T = \frac{1}{2} \left( e^{-f/2} + 2 e^{-f} + \gamma e^{-3 f/2} \right)\ .
\end{equation}
This result generalizes Bell's suggestion to $N_t = 2$ and already
reveals the characteristic structure of the solution for general
cluster size $N_t$: mean cluster lifetime $T$ is suppressed
exponentially by force and the rebinding correction is a polynomial
of power $\gamma^{N_t-1}$. A detailed analysis shows that although very
different for $f < f_c$, for $f > f_c$ the stochastic treatment gives
similar results in regard to $T$ as the deterministic one.

In order to investigate the effect of fluctuations, we used computer
simulations to numerically solve the master equation
\eq{MasterEquation}. This can in fact be done very efficiently
by using the Gillespie algorithm for exact stochastic simulations
\cite{c:gill76}. Our computer simulations show that for $f > f_c$, 
single rupture trajectories $i(\tau)$ show a characteristic shape
which is not revealed by considering the first moment $\langle i(\tau)
\rangle$ only. Initially they follow the average value, but 
then they abruptly move towards the completely dissociated state,
while the average value approaches this state in a more gentle way.
Therefore the average behaviour results not so much from differently
shaped trajectories, but rather from the distribution of the
timepoints of abrupt decay.  This observation shows the importance of
fluctuations and can be understood from the rates $r_i$ given in
\eq{Rates}: once there is a fluctuations to a smaller number of closed
bonds, force on the remaining bonds rises and leads to even more
increased dissociation. Therefore a positive feedback exists for bond
rupture, which for $f > f_c$ cannot be balanced anymore by rebinding
effects.

It is well known that bifurcations often lead to switch-like behaviour
in biochemical control systems \cite{s:tyso03}. In general, thresholds
have evolved for many biological systems, including the cell cycle and
the MAPK-cascade. Our model shows that switch-like behaviour can also
arise from the mechanical effect of force. Similar mechanisms are very
likely to be at work at focal adhesions. In particular, the
experimental evidence described above suggests that a certain
threshold of force is required to trigger signaling events which
eventually lead to regulated growth of focal adhesions. Since the
build-up of internal force has to be balanced by the extracellular
environment, its mechanical properties modulate the way in which
the threshold is reached. Therefore an internal threshold for force
is an appealing candidate for the exact mechanism
of the mechanosensor at focal adhesions.

\section{The two-spring model}

\begin{figure}
\begin{center}
  \includegraphics[width=0.7\textwidth]{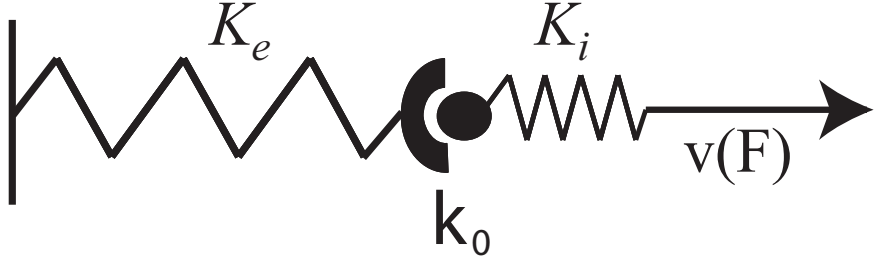}
\caption{In the two-spring model, the spring constant $K_e$ represents
extracellular elasticity and the spring constant $K_i$ represents the
mechanical properties of the intracellular structure.  Force
generation by the actin cytoskeleton is represented by the linearized
force-velocity relation $v(F)$ for a single myosin II molecular
motor. The internal state of the focal adhesion is represented by a
biomolecular bond which opens in a stochastic manner with dissociation
rate $k_0$.}
\label{TwoSpringsCartoon}
\end{center}
\end{figure}

\begin{figure}
\begin{center}
\includegraphics[width=0.7\textwidth]{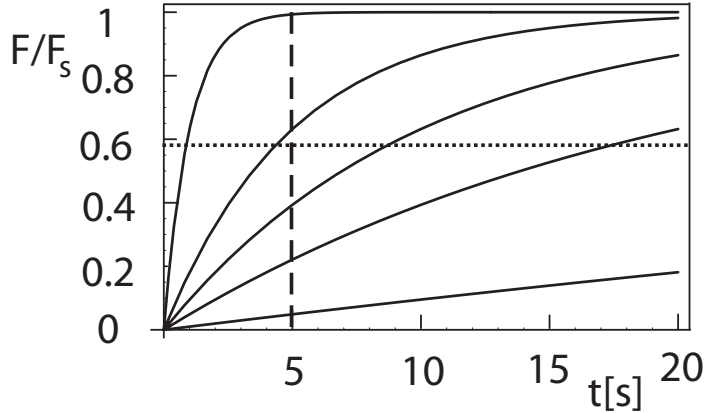}
\caption{Build-up of force from \eq{eq:Ft} as resulting 
from the two-spring model from \fig{TwoSpringsCartoon}. Free velocity
$v_0 = 10$ $\mu$m/s, stall force $F_s = 10$ pN and spring constant $K
= 1, 5, 10, 20, 100$ pN/$\mu$m (from bottom to top).  The horizontal
line marks some putative threshold in force which might be required to
activate the signaling which stabilizes focal adhesions. The vertical
line marks some hypothetical time scale which might characterize the
internal state of focal adhesions.}
\label{fig:Ft}
\end{center}
\end{figure}

In order to investigate this point in quantitative detail, we now
introduce a simple two-spring model for build-up of force at focal
adhesions. The model is depicted schematically in
\fig{TwoSpringsCartoon}. Here the ECM and the force-bearing
intracellular structures are represented by harmonic springs with
spring constants $K_e$ and $K_i$, respectively.  Since the two springs
act in series, the effective spring constant is given by $1 / K = 1 /
K_e + 1 / K_i$. Therefore the overall stiffness $K$ is mainly
determined by the softer spring. For the time being we assume that
this applies to the extracellular environment. Tension in the actin
stress fibers is generated by myosin II molecular motors. For
simplicity, we represent their activity by a linearized force-velocity
relation
\begin{equation} \label{ForceVelocityRelation}
v(F) = v_0 (1 - \frac{F}{F_s})
\end{equation}
where free velocity is of the order of $v_0 = 10$ $\mu$m/s and stall
force $F_s$ is a few pN \cite{b:howa01}.  As the motors pull, the
springs get strained. For the static situation, the energy $W = F^2 /
2 K$ is stored in the spring. Therefore the stiffer the environment,
the less work has to be invested into building up a certain level of
force $F$. For the dynamic situation, we have $dW = F dF / K$. The
dynamics of force generation can be derived by noting that the power
$dW / dt$ invested into the spring is generated by the molecular
motors:
\begin{equation} \label{PowerBalance}
\frac{dW}{dt} = \frac{F}{K} \frac{dF}{dt} = F v(F)
\end{equation}
with the force-velocity relation from \eq{ForceVelocityRelation}.
This equation can be readily integrated:
\begin{equation} \label{eq:Ft}
F = F_s \left( 1 - e^{-t / t_K} \right)
\end{equation}
with $t_K = F_s / v_0 K$. If the cell pulls on a material with a bulk
modulus of kPa, then the corresponding spring constant on the
molecular level can be expected to be of the order of $K = pN / \mu
m$.  Thus the typical time scale $t_K$ is seconds. If the bulk modulus
is of the order of MPa, then $K = pN / nm$ (which is also the range
for protein stiffness) and the typical time scale $t_K$ is
milliseconds. In \fig{fig:Ft}, we plot \eq{eq:Ft} for different values
of the spring constant $K$.  All curves eventually saturate at $F =
F_s$, but the stiffer the environment (the larger $K$), the faster a
given threshold in force (indicated by the horizontal dashed line) can
be reached.

Like the general adhesion clusters discussed in the preceding section,
focal adhesions are also subject to force-modulated internal
dynamics. In the two-spring model from \fig{TwoSpringsCartoon}, the
internal structure of the focal adhesion is represented by one
biomolecular bond with unstressed dissociation rate $k_0$. In
principle one now can apply the concept of rupture under force to the
loading history from \eq{eq:Ft}. In fact recent years have shown that
rupture under non-constant force is essential to understand the
details of biomolecular bonding \cite{c:evan01a}.  For linear ramps of
force, this issue has been addressed theoretically in great details,
both for single bonds \cite{c:evan97,c:seif98} and adhesion clusters
\cite{c:seif00,uss:erdm04b}.  Unfortunately, the differential equation
for the probability $p(t)$ that one bond with the loading history from
\eq{eq:Ft} breaks at time $t$ can be solved only
numerically. Therefore it is instructive to consider two simple
limites of this situation. As in the preceding section, we assume that
the single bond under constant force $F_s$ has the average lifetime $T
= e^{- F_s / F_b} / k_0$. In the case of large $K$, $t_K < T$ and the
bond effectively experiences constant loading with stall force $F_s$.
In the case of small $K$, loading is approximately linear, with a
loading rate $F_s / t_K$. If the dimensionless loading rate $F_s /
(t_K k_0 F_b) = (v_0 K) / (k_0 F_b) < 1$, then the bond will decay
with its intrinsic rate $k_0$ before the effect of force becomes
relevant. The general case will be within these two limites.  Since
the stall force $F_s$ is of the same order as the internal force scale
$F_b$, the effect of the loading history is expected to change the
result by not more than one order of magnitude. For simplicity, we
therefore now use the force-independent dissociation rate $k_0$. Then
we deal with a Poisson process with an exponentially decaying
probability $p(t) = e^{- k_0 t} k_0 dt$ that the bond breaks at time
$t$ in a time interval $dt$. Using \eq{eq:Ft}, we then calcuate the
average force which has been built up until bond rupture:
\begin{equation} \label{eq:average_force}
\langle F \rangle = \int_0^{\infty}p(t)F(t)dt=\frac{F_s}{1+k_0 t_K}\ .
\end{equation}
We therefore conclude that the level of force reached is essentially
determined by the quantity $k_0 t_K = (k_0 F_s) / (v_0 K)$. Since
unstressed dissociation constant $k_0$, stall force $F_s$ and maximal
motor velocity $v_0$ are molecular constants, the only relevant
quantity in this context is indeed the external stiffness $K$.  Using
the typical values given above, we find that $k_0 t_K$ is of the order
of $1$ and $10^{-3}$ for soft and stiff springs, respectively.  This
results in an average force $\langle F \rangle$ which is larger by a
factor of $2$ in the stiff environment. Note that this outcome for the
average force $\langle F \rangle$ is somehow weaker than one would
expect by naively inspecting \fig{fig:Ft} in regard to the level of
force $F$ reached after some internally determined time $1/k_0$
(indicated by the vertical dashed line).

The simple two-spring model can now be used to make first quantitative
predictions for active mechanosensing at focal adhesions. If a cell is
pulling at several focal adhesions with a similar investment of
resources, then those contacts will reach the level of force
putatively required for activation of the relevant signaling pathways
which experience the largest local stiffness in their environment.
Growth of contacts in an elastically anisotropic environment might
then lead to cell polarization and locomotion in the direction of
maximal effective stiffness in the environment, which has been
observed experimentally for different adhesion-dependent cell types on
elastic substrates \cite{c:lo00,c:wong03}. Recently we have shown that
such an effective cell behaviour can be described by an extremum
principle in linear elasticity theory \cite{uss:bisc03a,uss:bisc04a}.
Solving the elastic equations for different geometries and boundary
conditions of interest, one then can predict non-trivial effects for
cell positioning and orientation, in good agreement with numerous
experimental observations for cells on elastic substrates and in
hydrogels. The two-spring model introduced here also makes interesting
predictions regarding the way cells perceive extracellular
rigidity. Since $1 / K = 1 / K_e + 1 / K_i$, cells can only perceive
external stiffness relative to their internal stiffness.  This
suggests that cells have mechanisms to match their internal with the
external stiffness.

\section{Conclusions}

In order to understand mechanotransduction processes in animal tissues
and organs quantitatively and on the systems level, one has to
investigate the way the mechanical properties of the environment, the
regulation of actomyosin contractility and the conversion of physical
force into biochemical signals work together at focal adhesions. Here
we have presented first quantitative steps in this direction.  We
first discussed integrin signaling from focal adhesions and how it
feeds back to the integrins through the actin cytoskeleton.  Next we
discussed a model for the rupture dynamics of adhesion clusters under
force which showed that force is an important regulator of the
internal state of focal adhesions and that switch-like control
mechanisms can result from a structural model. Introducing the
two-spring model, we then showed how this internal dynamics can in
principle be coupled to extracellular elasticity and intracellular
force generation.  Our treatment shows how biochemical and structural
aspects might be coupled at focal adhesions. In order to arrive at a
complete and predictive understanding of focal adhesions, future work
has to develop new concepts along these lines and to incorporate as
much experimental data as possible. In the long run, this effort then
might become an important part of the future systems biology of
tissues and organs.

\textit{Acknowlegdments:} This work was supported by the Emmy Noether
Program of the German Research Foundation (DFG) and the Center for
Modelling and Simulation in the Biosciences (BIOMS) at Heidelberg
University.


\end{document}